\documentclass[fleqn,usenatbib]{mnras}
\usepackage{newtxtext,newtxmath}
\usepackage[T1]{fontenc}
\DeclareRobustCommand{\VAN}[3]{#2}
\let\VANthebibliography\thebibliography
\def\thebibliography{\DeclareRobustCommand{\VAN}[3]{##3}\VANthebibliography}

\usepackage{graphicx}	% Including figure files
\usepackage{amsmath}	% Advanced maths commands

\usepackage{amssymb}	% Extra maths symbols

\usepackage{soul}

\newcommand{\HI}{H{\sc i}}
\newcommand{\eg}{{\emph{e.g.}}}
\newcommand{\ie}{{\emph{i.e.}}}
\newcommand{\kmps}{$\,\rm{km} \, \rm{s}^{-1}$\,}
\defcitealias{Yang2007}{Y07}
\defcitealias{Saulder2016}{S16}
\title[Scatter of the H{\sc i}-halo mass relation of centrals]{xGASS: The scatter of the H{\sc i}-halo mass relation of central galaxies}

\author[M. Saraf et al.]{
Manasvee Saraf,$^{1,2,3}$\thanks{E-mail: manasvee.saraf@icrar.org}
Luca Cortese,$^{1,3}$
O. Ivy Wong,$^{2,1}$
Barbara Catinella,$^{1,3}$
Steven Janowiecki,$^{4}$ and
\newauthor Jennifer A. Hardwick.$^{1,3}$
\\
$^{1}$International Centre for Radio Astronomy Research, The University of Western Australia, Crawley, WA 6009, Australia\\
$^{2}$Australia Telescope National Facility, CSIRO, Space and Astronomy, P.O. Box 1130, Bentley, WA 6102, Australia\\
$^{3}$ARC Centre of Excellence for All-Sky Astrophysics in 3 Dimensions (ASTRO 3D), Australia\\
$^{4}$University of Texas, Hobby-Eberly Telescope, McDonald Observatory, TX 79734, USA\\
}

\date{Accepted XXX. Received YYY; in original form ZZZ}

\pubyear{2023}

\begin{document}
\label{firstpage}
\pagerange{\pageref{firstpage}--\pageref{lastpage}}
\maketitle
\begin{abstract}
Empirical studies of the relationship between baryonic matter in galaxies and the gravitational potential of their host halos are important to constrain our theoretical framework for galaxy formation and evolution.
One such relation, between the atomic hydrogen (\HI) mass of central galaxies ($M_{\rm{HI,c}}$) and the total mass of their host halos ($M_{\rm{halo}}$), has attracted significant interest in the last few years.
In this work, we use the extended GALEX Arecibo SDSS Survey to examine the scatter of the \HI-halo mass relation for a representative sample of central galaxies.
Our findings reveal a flat median relation at $\rm{log}_{10}$$(M_{\rm{HI,c}}/\rm{M}_{\odot}) \approx 9.40$, across $11.1 < \rm{log}_{10}$$(M_{\rm{halo}}/\rm{M_{\odot}}) < 14.1$. 
This flat relation stems from the statistical dominance of star-forming, disc galaxies at low $M_{\rm{halo}}$ in combination with the increasing prevalence of passive, high stellar-concentration systems at higher $M_{\rm{halo}}$.
The scatter of this relation and the stellar specific angular momentum of centrals have a strong link (Spearman's rank correlation coefficient $\geq 0.5$).
Comparisons with simulations suggest that the kinematic state of host halos may be primarily driving this scatter.
Our findings highlight that the \HI-halo mass parameter space is too complex to be completely represented by simple median or average relations and we show that tensions with previous works are most likely due to selection biases.
We recommend that future observational studies, and their comparisons with theoretical models, bin central galaxies also by their secondary properties to enable a statistically robust understanding of the processes regulating the cold gas content within central galaxies of dark-matter halos.
\end{abstract}


%%%%%%%%%%%%%%%%% BODY OF PAPER %%%%%%%%%%%%%%%%%%

\section{Introduction}
\label{sec:Start}
Studies that link the baryonic content of galaxies to the gravitational field in and around them, are important to further the understanding of the nature of dark matter and the processes that govern galaxy formation and evolution.
Cold gas plays a critical role in the process of star formation and, thus, is a probe for processes regulating galaxy evolution.
However, the relationship between cold gas and the gravitational potential of its host halo has not been completely understood.
In the last few years, the atomic hydrogen (\HI) to halo mass relation has received increasing attention, in particular from the theoretical community, as it can help constrain simulations of galaxy formation \citep[\eg][]{Kim2017a,Padmanabhan2017, Villaescusa-Navarro2018, Baugh2019, Spinelli2020, Chauhan2020}.
Currently, different simulations predict different values of \HI\ mass ($M_{\rm{HI}}$) for a given host halo mass ($M_{\rm{halo}}$), particularly in the $11.5 < \rm{log}_{10}$$(M_{\rm{halo}}/\rm{M_{\odot}}) < 13.5$ range.
These disagreements arise due to different input physical models such as that of gas dynamics, stellar feedback, ram-pressure effects, photo-ionisation and interstellar medium properties.

Simulations also predict a large scatter of the \HI-halo mass relation \citep{Villaescusa-Navarro2018,Baugh2019,Spinelli2020,Chauhan2020}.
These simulation-based studies have tested several physical drivers of this scatter such as that related to halo assembly history (\eg\ spin, substructure, concentration, formation age) and the efficiency of suppressing gas cooling and enhancing gas outflow by the active galactic nucleus (AGN; \eg\ black hole-to-stellar mass ratio of the central galaxy).
However, like the shape of this relation, the scatter too is found to be model dependent.
Testing these predictions with observations is challenging due to limitations of current observational data.
It is still very difficult to detect cold gas in galaxies below the star-forming main-sequence (SFMS), given the current sensitivity of radio telescopes \citep{Saintonge2022}.
To combat this bias and enhance the number statistics, several observational studies improve the signal-to-noise of the \HI\ 21 cm profile by co-adding spectra of optically selected galaxies or groups \citep[\eg][]{Guo2020, Rhee2023, Dev2023}.
These `spectral stacking' techniques provide direct measurements, extracted from fixed spatial apertures and line widths, of the average \HI\ content in galaxies or groups within selected $M_{\rm{halo}}$ bins.
However, while insightful in studying the underlying average trend between $M_{\rm{HI}}$ and $M_{\rm{halo}}$, these techniques do not provide any information regarding the scatter.

In simulation-based studies, the scatter of the \HI-halo mass relation is especially large when considering only the \HI\ mass within centrals ($M_{\rm{HI,c}}$) as opposed to that within the entire halo \citep[for example, see Fig. 7 and 10 of][]{Chauhan2020}.
Characterising this large scatter of centrals with observations could provide important constraints for the physical models input into various galaxy formation simulations.
In the past year, a few studies \citep{Dutta2022,Korsaga2023,Hutchens2023} have surfaced that show the observed $M_{\rm{HI}}-M_{\rm{halo}}$ relation to have a very small scatter, at odds with what is found using simulations.
This disagreement brings into question whether observation-based results are still biased due to selection effects, or simulation-based studies are using physical models that do not encapsulate the full complexity of baryon physics.
To answer these questions, we investigate the scatter of the \HI-halo mass relation of centrals from the extended GALEX Arecibo SDSS survey \citep[xGASS,][]{Catinella2018}.
For the first time, we extend previous works into the gas-poor regime, using a sample that provides a fair representation of the range of $M_{\rm{HI,c}}$ that is observed in local central galaxies residing in halos of a wide mass range, $11.1 < \rm{log}_{10}$$(M_{\rm{halo}}/\rm{M_{\odot}}) < 14.1$.

This paper is structured as follows.
Section \ref{sec:Data} describes the xGASS $M_{\rm{HI}}$ catalogue, two group catalogues using different $M_{\rm{halo}}$ estimation techniques and the studies measuring secondary galaxy properties.
The analysis on the shape and scatter of the \HI-halo mass relation of xGASS centrals is presented in section \ref{sec:Results}, where we also investigate the physical drivers of a central's position in this parameter space.
In section \ref{sec:Discussion}, we compare our results to those found using previous observation- and simulation-based studies before summarising our study and its key outcomes in section \ref{sec:End}.
Throughout this study, we assume a $\Lambda$ cold dark matter cosmology with $\Omega_{\rm{M}} = 0.3$, $\Omega_{\Lambda} = 0.7$ and the dimensionless Hubble parameter, $h=0.7$ related to the Hubble constant, $H_{0} = 70$ \kmps Mpc$^{-1}$.

\section{Data and Methodology}
\label{sec:Data}
\subsection{xGASS}
In this work, we use $M_{\rm{HI}}$ measurements, either detections or upper limits, from xGASS.
This survey of 1179 galaxies is selected from the overlap between the Sloan Digital Sky Survey Data Release 7 \citep[SDSS DR7,][]{Abazajian2009} spectroscopic catalogue and the GALaxy Evolution eXplorer \citep[GALEX,][]{Martin2005} sky footprint.
The selection of these xGASS galaxies is only based on stellar mass, $9 < \rm{log}_{10}$$(M_{*}/\rm{M}_{\odot}) < 11.5$, and redshift, $0.01 < z < 0.05$.
Moreover, the stellar mass ($M_{*}$) distribution of the observed galaxies is intentionally flat to improve statistics at the high $M_{*}$ end \citep[see Fig. 3 of][]{Catinella2018}. 
The galaxies were observed with Arecibo, a 305m single-dish radio telescope, until \HI\ was detected or an \HI\ gas-fraction limit of $\sim 2 \%$ for galaxies with $\rm{log}_{10}$$(M_{*}/\rm{M}_{\odot}) > 9.7$ or a limit of $\rm{log}_{10}$$(M_{\rm{HI}}/\rm{M}_{\odot}) = 8$ for galaxies with $\rm{log}_{10}$$(M_{*}/\rm{M}_{\odot}) \leq 9.7$ was achieved.
Thus, xGASS is largely gas-fraction limited and representative of not only \HI-rich but also \HI-poor galaxies.
We use this survey as it is the most sensitive and representative \HI\ survey of the local Universe and, therefore, is ideal to study the positions of individual galaxies on the \HI-halo mass parameter space.
The \texttt{HIsrc} flag in the xGASS catalogue is used to distinguish whether a galaxy has an $M_{\rm{HI}}$ detection or an upper limit.
For \HI-undetected galaxies, the $M_{\rm{HI}}$ upper limits are estimated by assuming a flat flux of 5 times the spectral noise across a velocity width of 200 \kmps or 300 \kmps depending on $M_{*}$ \citep[for details see][]{Catinella2018}.

\subsection{Group catalogues}
\label{sec:GroupCat}

For 1155 of the catalogued galaxies, the xGASS team provides environment metrics from the \citet[][hereafter \citetalias{Yang2007}]{Yang2007} group `B' catalogue \citep[for details see][]{Janowiecki2017}.
\citetalias{Yang2007} produced the group catalogue using a halo-based group finder on spectroscopic redshifts, mainly acquired from SDSS DR7.
From this \citetalias{Yang2007} group catalogue, \citet{Janowiecki2017} removed smaller counterparts of a shredded galaxy on the basis of the galaxies' separation from each other.
After matching to this shredding-corrected group catalogue, an environment code (\texttt{env\_code\_B}), the total number of galaxy members in the host group ($N_{\rm{gal}}$) and, where available, $M_{\rm{halo}}$ values are provided for each xGASS galaxy.
24 xGASS galaxies do not have matches with the \citetalias{Yang2007} group catalogue due to their close proximity to either bright stars or survey edges.
In \citetalias{Yang2007}, $M_{\rm{halo}}$ values are derived using an abundance-matching technique based on $M_{*}$ and only provided for massive halos with $\rm{log}_{10}$$(M_{\rm{halo}}/\rm{M}_{\odot}) \geq 11.5$.
For 343 xGASS centrals in less massive halos, we estimate $M_{\rm{halo}}$ by using a conditional stellar mass function \citep[CSMF; see model 6 in the Table 3 of][]{Yang2012}, which gives the stellar mass of a central ($M_{\rm{*,c}}$) as a function of its host's $M_{\rm{halo}}$.
We correct this CSMF to the Hubble parameter used by xGASS, $h = 0.7$, and translate it to fit the 197 group centrals with catalogued $M_{\rm{*}}$ and abundance-matching $M_{\rm{halo}}$ values.
We run this CSMF, given by equation \ref{eq:mHaloEst}, as an estimator to find the value of $M_{\rm{halo}}$ that returns the $M_{\rm{*,c}}$ of a given central.
\begin{equation} 
    \frac{M_{\rm{*,c}}}{\rm{M}_{\odot}} = 10^{10.22} \times \frac{\left(\frac{\frac{M_{\rm{halo}}}{\rm{M}_{\odot}}-10^{0.15}}{10^{10.78}}\right)^{7.94}}{\left( 1 + \frac{\frac{M_{\rm{halo}}}{\rm{M}_{\odot}}-10^{0.15}}{10^{10.78}}\right)^{7.66}} + 10^{0.05},
    \label{eq:mHaloEst}
\end{equation}

There are various uncertainties involved in the estimation of $M_{\rm{halo}}$ based on the method used by \citetalias{Yang2007}.
Thus, to confirm the results of our study we also use dynamical $M_{\rm{halo}}$ estimates from the \citet[][hereafter \citetalias{Saulder2016}]{Saulder2016} SDSS Data Release 12 (DR12) group catalogue.
\citetalias{Saulder2016} produced the group catalogue using a friends-of-friends based group finder that is optimised to account for observational biases in SDSS using mock catalogues from the Millennium simulation \citep{Springel2005}.
The dynamical $M_{\rm{halo}}$ is derived for each group depending on $N_{\rm{gal}}$.
For galaxies in groups with $N_{\rm{gal}} \geq 5$, $M_{\rm{halo}}$ is derived from a function of the group's total luminosity ($L_{\rm{tot}}$), luminosity distance ($D_{\rm{L}}$), velocity dispersion ($\sigma_{\rm{group}}$), radius ($R_{\rm{group}}$) and member galaxies detected in SDSS DR12 ($N_{\rm{fof}}$).
For smaller groups with $2 \leq N_{\rm{gal}} \leq 4$, $M_{\rm{halo}}$ is derived from a function of $L_{\rm{tot}}$, $D_{\rm{L}}$, $\sigma_{\rm{group}}$ and $R_{\rm{group}}$.
For isolated galaxies, $M_{\rm{halo}}$ is derived from a function of $L_{\rm{tot}}$ and $D_{\rm{L}}$ alone \citepalias[see section 3.4.7 of][for detailed equations and descriptions]{Saulder2016}. 
We assign an environment code (\texttt{rank\_Mstar}) to all galaxies in the \citetalias{Saulder2016} group catalogue by ranking those with the same group ID according to their $M_{*}$ value.
We define a central as the galaxy with the highest $M_{*}$ within a group.
Next, we perform a positional cross-match between the xGASS catalogue and the galaxy list of the group catalogue \citepalias[Table A.7. of][]{Saulder2016}.
Since both xGASS and the \citetalias{Saulder2016} group catalogue use right ascension and declination values from SDSS, we set a small 3 arcsec match radius.
We acquire 1160 matched galaxies and find that the 19 unmatched galaxies are in close proximity to bright stars or survey edges.
To recover the $M_{\rm{halo}}$ and $N_{\rm{gal}}$ group properties, we match the group IDs between the group list and the galaxy list of the group catalogue \citepalias[Table A.2. and A.7. of][respectively]{Saulder2016}.
Isolated galaxies are defined as the ones in groups with $N_{\rm{gal}} = 1$.

\subsection{Galaxy properties}
We use the following galaxy properties to study the drivers of the scatter of the \HI-halo mass relation.

\textbf{Stellar mass: } 
The xGASS catalogue provides $M_{*}$ estimates for every galaxy.
These $M_{*}$ values are obtained from the value-added SDSS DR7 catalogue of the Max Planck Institute for Astrophysics/Johns Hopkins University, assuming the initial mass function from \citet{Chabrier2003}.

\textbf{Star-formation rate: }
Star-formation rate (SFR) estimates are provided for 1171 galaxies in the xGASS catalogue.
SFR is calculated by combining near-ultraviolet (NUV) magnitudes from GALEX and 22 $\mu$m and/or 12 $\mu$m mid-infrared (MIR) magnitudes measured from the Wide-field Infrared Survey Explorer \citep[WISE,][]{Wright2010} images.
Where NUV or MIR fluxes are not available, SFR is estimated using spectral energy distribution fits from \citet{Wang2011} \citep[for details see][]{Janowiecki2017}.
We calculate $\Delta\rm{MS}$ (deviations from the SFMS) to acquire a quantification of the star-formation activity that is independent of $M_{*}$.
For the subset of xGASS galaxies with catalogued SFR values, $\Delta\rm{MS}$ is estimated by calculating specific SFR and using the xGASS SFMS fit \citep[equation 2 of][]{Catinella2018}.
We use a threshold of $\Delta\rm{MS} = -0.5$ to separate star-forming galaxies ($\Delta\rm{MS} > -0.5$) from galaxies below the SFMS ($\Delta\rm{MS} \leq -0.5$, or passive).

\textbf{Stellar-concentration index: } 
The xGASS catalogue provides stellar-concentration indices ($C_{i}$) for every galaxy.
$C_{i}$ is calculated by dividing the radius containing 90\% of the flux, $R_{90}$, by the radius containing 50\% of the flux, $R_{50}$.
$R_{90}$ and $R_{50}$ are measured in arcsec, provided in the SDSS DR7 database and based on Petrosian fluxes in the $r$ band.
We define the galaxies with $C_{i} < 2.6$ as `disc-dominated systems' and those with $C_{i} \geq 2.6$ as `high stellar-concentration systems'.

\textbf{Stellar specific angular momentum: } 
\citet{Hardwick2022} provide stellar specific angular momentum ($j_{*}$) estimates for 564 xGASS \HI-detected galaxies of inclinations, $i > 30^{\circ}$.
$j_{*}$ is calculated within an aperture of 10 times the $r-$band half-light radius of the disc \citep[provided by][]{Cook2019}.
These calculations assume that the galaxy's rotational velocity is constant and set by the \HI\ profile width, therefore, assuming that the stars are co-rotating with the \HI\ and the bulge is co-rotating with the disc.
Any assumptions made do not affect the $j_{*}$ estimates by more than 0.1 dex \citep[for details, refer to][]{Hardwick2022}.

\subsection{The final sample}
\label{sec:centralselection}
Before the last step of our sample selection, we remove the `confused' galaxies that are affected by the presence of more dominant sources of \HI\ emission within the same $\sim 3.5$ arcmin Arecibo beam.
For \HI-detected galaxies (\ie\ those with the xGASS flag, \texttt{HIsrc} < 4), confused galaxies are indicated in the xGASS catalogue by an \HI\ quality flag of \texttt{HI\_FLAG} $\geq 3$.
We first remove these confused galaxies and then select centrals using the environment code (\texttt{env\_code\_B} > 0 or \texttt{rank\_Mstar} > 0).
For \HI-undetected galaxies, \texttt{HI\_FLAG} is not provided and centrals are selected based on their environment code alone.
Table \ref{tab:counts} summarises the statistics of our final sample of centrals based on their environmental and \HI-detection properties.
Table \ref{tab:counts} also lists the statistics in the `$\Delta\rm{MS}$ sub-samples' (\ie\ xGASS centrals with SFR estimates) and the `$j_{*}$ sub-samples' (\ie\ xGASS centrals with $j_{*}$ estimates).
In this paper, `centrals' refers to the combination of isolated galaxies and group centrals.

\begin{table*}
\centering
\caption{Statistics of xGASS centrals, matched to the \citetalias{Yang2007} and \citetalias{Saulder2016} group catalogues, categorised by their environmental and \HI\ detection properties. Additional columns refer to centrals with catalogued $\Delta\rm{MS}$ and $j_{*}$ estimates. Filters are applied to ensure that all the \HI-detected centrals are not affected by beam confusion.}
\label{tab:counts}
\begin{tabular}{ll|c|c|c|c|c|c|}
 &   & \multicolumn{3}{|c|}{\citetalias{Yang2007}} & \multicolumn{3}{|c|}{\citetalias{Saulder2016}} \\
 &   & full sample & $\Delta\rm{MS}$ sub-sample & $j_{*}$ sub-sample & full sample & $\Delta\rm{MS}$ sub-sample & $j_{*}$ sub-sample \\ \hline\hline
\multicolumn{1}{|l|}{} & all measurements & 531 & 528 & 344 & 369 & 366 &  251 \\ \cline{2-8}
\multicolumn{1}{|l|}{isolated galaxies} & \HI\ detection & 415 & 413 & 344 & 300 & 298 &  251 \\ \cline{2-8}
\multicolumn{1}{|l|}{} & \HI\ upper limit & 116 & 115 & - & 69 & 68 & - \\ \hline\hline
\multicolumn{1}{|l|}{} & all measurements & 233 & 232 & 83 & 285 & 284 & 111 \\ \cline{2-8}
\multicolumn{1}{|l|}{group centrals} & \HI\ detection & 137 & 137 & 83 & 165 & 165 & 111 \\ \cline{2-8}
\multicolumn{1}{|l|}{} & \HI\ upper limit & 96 & 95 & - & 120 & 119 & - \\ \hline\hline
\multicolumn{1}{|l|}{} & all measurements & 764 & 760 & 427 & 654 & 650 & 362 \\ \cline{2-8}
\multicolumn{1}{|l|}{total centrals} & \HI\ detection & 552 & 550 & 427 & 465 & 463 & 362 \\ \cline{2-8}
\multicolumn{1}{|l|}{} & \HI\ upper limit & 212 & 210 & - & 189 & 187 & - \\ \hline
\end{tabular}
\end{table*}

\section{Results}
\label{sec:Results}
We present the results of our study in this section.
The strength of the correlation between two parameters is quantified by calculating the Spearman's rank correlation coefficient ($\rho_{\rm{S}}$), which is only considered significant and stated if it has a $p$-value $< 0.03$.

\subsection{The median \HI-halo mass relation of xGASS centrals}
\label{sec:HIHM}

\begin{figure*}
	\includegraphics[width=\textwidth]{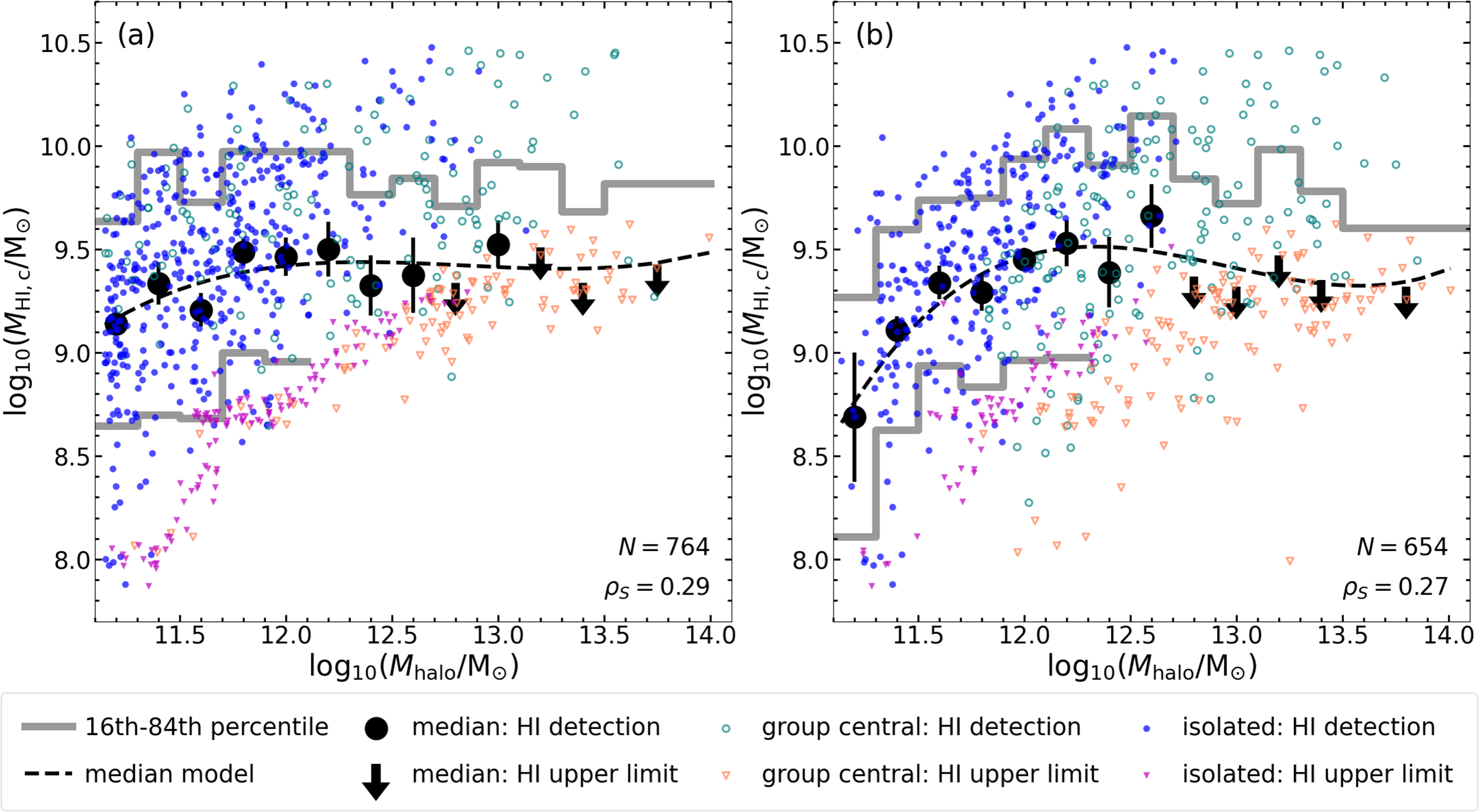}
    \caption{The \HI-halo mass relation of xGASS centrals, matched to the \citetalias{Yang2007} (panel a) and \citetalias{Saulder2016} (panel b) group catalogues.
    Isolated galaxies are marked with filled symbols and group centrals are marked with open symbols.
    \HI-detected isolated/group centrals are displayed as blue/cyan circles, and those with \HI\ mass upper limits are displayed as magenta/orange downward-facing arrows.
    The median \HI\ mass values are displayed in black and marked by circles with error-bars, if consistent with \HI-detected centrals, or by downward-facing arrows, if consistent with \HI-undetected centrals.
    Dashed black lines mark the median model \HI-halo mass relations given by equation \ref{eq:HIHMmodel_Yang} (panel a) or equation \ref{eq:HIHMmodel_Saulder} (panel b).
    In each bin, the region between the grey lines shows the 16th-84th percentile of the \HI\ mass distribution, marking a large \HI\ mass scatter at all halo masses.
    The 16th percentile is not constrained at the high halo mass end, where \HI\ mass upper limits dominate.}
    \label{fig:HIHM}
\end{figure*}

Fig. \ref{fig:HIHM} a and b display the \HI-halo mass parameter space with the xGASS centrals matched to the \citetalias{Yang2007} and \citetalias{Saulder2016} group catalogues respectively.
The points are categorised by the environmental and \HI\ detection properties of the centrals.
It is clear that our sample is dominated by \HI-detected, isolated galaxies where $\rm{log}_{10}$$(M_{\rm{halo}}/\rm{M}_{\odot}) < 12.5 (12.3)$ and \HI-undetected, group centrals where $\rm{log}_{10}$$(M_{\rm{halo}}/\rm{M}_{\odot}) \geq 12.7 (12.9)$ in Fig. \ref{fig:HIHM}a (Fig. \ref{fig:HIHM}b).
We note that the sample in Fig. \ref{fig:HIHM}a, as opposed to that in Fig. \ref{fig:HIHM}b, has a higher fraction of isolated galaxies, \ie\ 70\% and 56\% respectively.

To estimate the median \HI-halo mass relation and quantify the variance of its scatter, we bin our sample of centrals by the $M_{\rm{halo}}$ values of their host halos, and calculate the median $M_{\rm{HI,c}}$ for each bin.
Throughout this paper, we use fixed bin widths of 0.2 dex.
However, if the counts are $< 10$, we set a wider bin width to ensure good statistics. 
For each $M_{\rm{halo}}$ bin, Table \ref{tab:HIHM} lists the $M_{\rm{halo}}$ limits, galaxy counts ($N$) and the median $M_{\rm{HI,c}}$ value. 
We prefer to use medians (rather than averages) because these are not affected by the non-detections, as long as the median value itself is a detection. 
Table \ref{tab:HIHM} also provides the scatter, quantified as the 16th-84th percentile of the $M_{\rm{HI,c}}$ distribution in that bin.
The uncertainties of the median $M_{\rm{HI,c}}$ values are determined via bootstrapping, generating 100,000 random samples for each bin of halo mass.
Note, that if the median $M_{\rm{HI,c}}$ value is consistent \HI-undetected centrals, we show it as an upper limit. 
Similarly, to highlight the parameter space that is unconstrained by our data, the percentile values are only shown if the percentile is consistent with \HI-detected centrals.
Median \HI-halo mass relations are modelled by fitting a third order polynomial to the mid $M_{\rm{halo}}$ and median $M_{\rm{HI,c}}$ values of each bin.
This median model is given by,
\begin{equation} 
    M_{\rm{HI,c}} = 0.08 M_{\rm{halo}}^{3} - 3.00 M_{\rm{halo}}^{2} + 38.41 M_{\rm{halo}} - 154.50,
    \label{eq:HIHMmodel_Yang}
\end{equation}
for the centrals matched to the \citetalias{Yang2007} group catalogue and by,
\begin{equation} 
    M_{\rm{HI,c}} = 0.20 M_{\rm{halo}}^{3} - 7.59 M_{\rm{halo}}^{2} + 98.10 M_{\rm{halo}} - 412.29,
    \label{eq:HIHMmodel_Saulder}
\end{equation}
for the centrals matched to the \citetalias{Saulder2016} group catalogue.

We find consistent medians and scatter in Fig. \ref{fig:HIHM} a and b, despite the different $M_{\rm{halo}}$ estimation methods used.
For $\rm{log}_{10}$$(M_{\rm{halo}}/\rm{M}_{\odot}) < 12.7$, there is a large scatter of $M_{\rm{HI,c}}$ values between 0.8-1.3 dex and the medians appear approximately constant at $\rm{log}_{10}$$(M_{\rm{HI,c}}/\rm{M}_{\odot}) \approx 9.40$.
At higher $M_{\rm{halo}}$ values, the medians are dominated by \HI-undetected centrals, meaning that the scatter of the relation is either similar to that observed at lower $M_{\rm{halo}}$ values or, most likely, larger.
In the next subsection we explore if there is a physical reason behind such a significant scatter of the relation.

\begin{table}
\centering
\caption{The median \HI\ mass values, marked in Fig. \ref{fig:HIHM} a and b, are presented alongside the halo mass limits and counts of centrals ($N$) for each bin. The 16th and 84th percentiles of the scatter in the \HI\ mass distribution of each bin is also provided where constrained.}
\label{tab:HIHM}
\begin{tabular}{crrr}
Bin & \multicolumn{1}{c}{$N$} & \multicolumn{1}{c}{Median} & \multicolumn{1}{c}{Scatter} \\
log$_{10}$$(M_{\rm{halo}})$  & & \multicolumn{1}{c}{log$_{10}$$(M_{\rm{HI,c}})$}  & \multicolumn{1}{c}{log$_{10}$$(M_{\rm{HI,c}})$}  \\ 
{[}M$_{\odot}${]} & & \multicolumn{1}{c}{{[}M$_{\odot}${]}} & \multicolumn{1}{c}{{[}M$_{\odot}${]}} \\ \hline \hline
\multicolumn{4}{c}{xGASS centrals matched to the \citetalias{Yang2007} group catalogue (Fig. \ref{fig:HIHM}a)} \\ \hline
{[}11.1, 11.3) & 106 & $9.14 \pm 0.06$ & {[}8.65, 9.64{]} \\
{[}11.3, 11.5) &  90 & $9.33 \pm 0.10$ & {[}8.70, 9.97{]} \\
{[}11.5, 11.7) & 116 & $9.21 \pm 0.08$ & {[}8.68, 9.73{]} \\
{[}11.7, 11.9) & 108 & $9.49 \pm 0.06$ & {[}9.00, 9.97{]} \\ 
{[}11.9, 12.1) &  88 & $9.46 \pm 0.09$ & {[}8.96, 9.97{]} \\
{[}12.1, 12.3) &  54 & $9.50 \pm 0.13$ & {[}-, 9.97{]} \\
{[}12.3, 12.5) &  44 & $9.33 \pm 0.15$ & {[}-, 9.76{]} \\ 
{[}12.5, 12.7) &  46 & $9.37 \pm 0.18$ & {[}-, 9.84{]} \\ 
{[}12.7, 12.9) &  43 &     $\leq 9.34$ & {[}-, 9.71{]} \\ 
{[}12.9, 13.1) &  27 & $9.52 \pm 0.11$ & {[}-, 9.92{]} \\ 
{[}13.1, 13.3) &  15 &     $\leq 9.51$ & {[}-, 9.90{]} \\ 
{[}13.3, 13.5) &  13 &     $\leq 9.34$ & {[}-, 9.68{]} \\ 
{[}13.5, 14.0) &  14 &     $\leq 9.43$ & {[}-, 9.82{]} \\ \hline
\multicolumn{4}{c}{xGASS centrals matched to the \citetalias{Saulder2016} group catalogue (Fig. \ref{fig:HIHM}b)} \\ \hline
{[}11.1, 11.3) &  15 & $8.69 \pm 0.31$ & {[}8.11, 9.27{]} \\
{[}11.3, 11.5) &  59 & $9.11 \pm 0.06$ & {[}8.62, 9.60{]} \\
{[}11.5, 11.7) &  63 & $9.34 \pm 0.08$ & {[}8.94, 9.74{]} \\
{[}11.7, 11.9) &  86 & $9.29 \pm 0.09$ & {[}8.83, 9.75{]} \\ 
{[}11.9, 12.1) &  90 & $9.45 \pm 0.06$ & {[}8.97, 9.94{]} \\
{[}12.1, 12.3) &  77 & $9.53 \pm 0.11$ & {[}8.98, 10.08{]} \\
{[}12.3, 12.5) &  69 & $9.39 \pm 0.17$ & {[}-, 9.91{]} \\ 
{[}12.5, 12.7) &  47 & $9.66 \pm 0.15$ & {[}-, 10.15{]} \\ 
{[}12.7, 12.9) &  39 &     $\leq 9.37$ & {[}-, 9.84{]} \\ 
{[}12.9, 13.1) &  35 &     $\leq 9.32$ & {[}-, 9.72{]} \\ 
{[}13.1, 13.3) &  25 &     $\leq 9.47$ & {[}-, 9.98{]} \\ 
{[}13.3, 13.5) &  28 &     $\leq 9.35$ & {[}-, 9.78{]} \\ 
{[}13.5, 14.1) &  21 &     $\leq 9.32$ & {[}-, 9.60{]} \\ \hline
\end{tabular}
\end{table}

\subsection{The scatter along the median \HI-halo mass relation}

\begin{figure*}
	\includegraphics[width=0.9\textwidth]{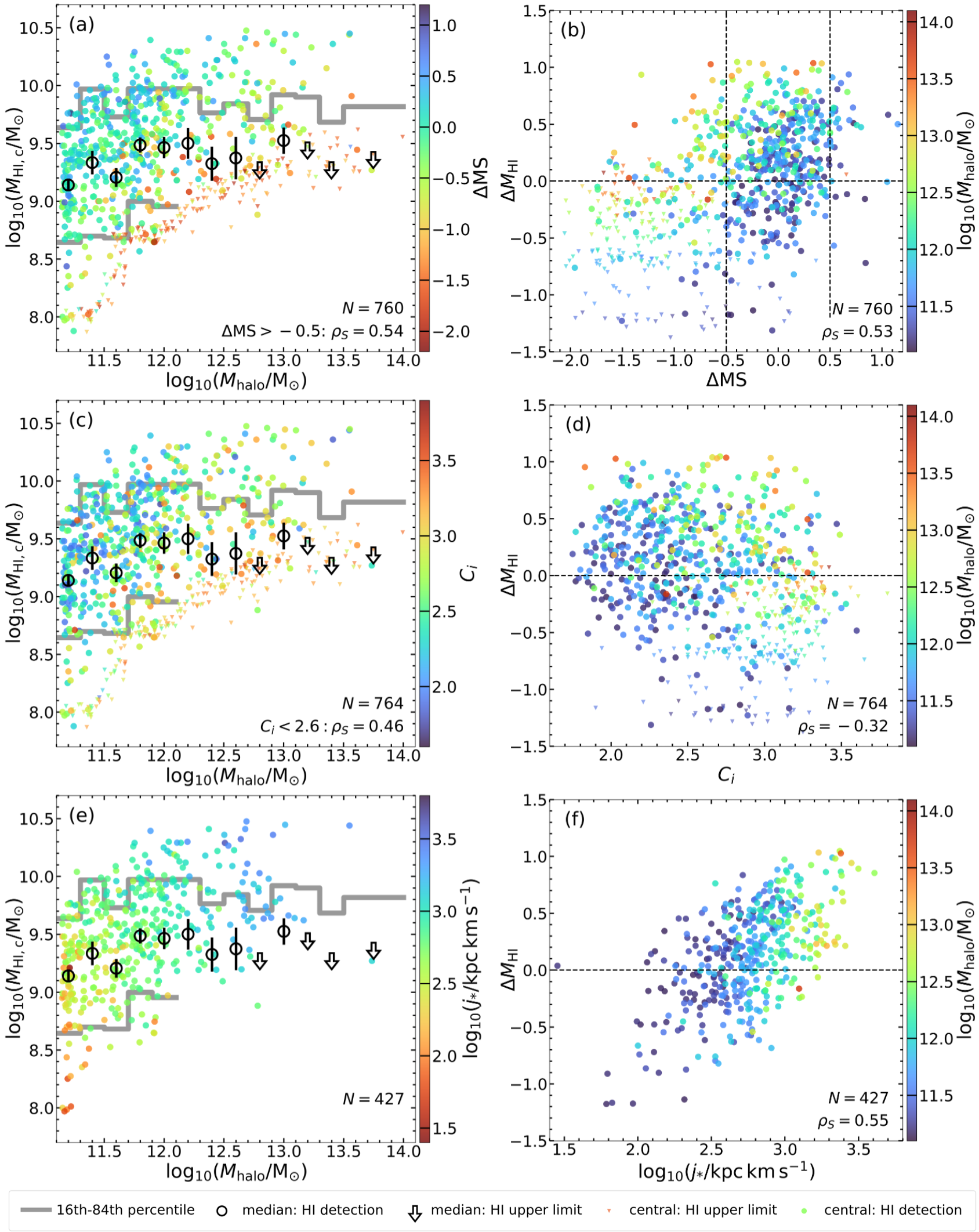}
    \caption{The scatter along the median \HI-halo mass relation of xGASS centrals, matched to the \citetalias{Yang2007} group catalogue.
    \HI-detected centrals are displayed as circles, and those with $M_{\rm{HI,c}}$ upper limits are displayed as downward-facing arrows.
    \textit{Left:} The \HI-halo mass parameter space, with points colour-coded by either $\Delta\rm{MS}$ (panel a), $C_{i}$ (panel c) or $j_{*}$ (panel e). 
    The median \HI\ mass values (open black symbols) and the 16th-84th percentile (region between grey lines) are displayed as in Fig. \ref{fig:HIHM}a.
    The $\rho_{\rm{S}}$ value between $M_{\rm{HI,c}}$ and $M_{\rm{halo}}$ is stated for star-forming and disc-dominated centrals in panels a and c respectively.
    \textit{Right:} $\Delta M_{\rm{HI}}$ as a function of either $\Delta\rm{MS}$ (panel b), $C_{i}$ (panel d) or $j_{*}$ (panel f), with their $\rho_{\rm{S}}$ values stated and points colour-coded by $M_{\rm{halo}}$.
    $\Delta M_{\rm{HI}}$ is calculated from the median model \HI-halo mass relation (equation \ref{eq:HIHMmodel_Yang}), which is represented by the horizontal dashed black line at $\Delta M_{\rm{HI}} = 0$.
    The SFMS is marked by the region between the two vertical dashed black lines in panel b.
    In each row, the limits of the colour bars in the left column are the same as the limits of the $x$-axis in the right column and vice versa.
    Note, the second row displays only the $\Delta\rm{MS}$ sub-sample and the bottom row displays only the $j_{*}$ sub-sample.}
    \label{fig:HIHMscatter_Yang}
\end{figure*}

In this section, we investigate which galaxy property correlates with the scatter of the \HI-halo mass relation.
Stellar, kinematic and environmental properties of a galaxy are known to correlate with its cold-gas properties \citep[\eg][]{Oosterloo2010, Odekon2016, Janowiecki2017, Stevens2019, Janowiecki2020, Li2022, Saintonge2022, Cortese2021}. 
To identify physical properties that might be driving the scatter, we investigated all the observables available for the xGASS sample. 
These include $N_{\rm{gal}}$, $M_{*}$, SFR, specific SFR, $\Delta\rm{MS}$, $C_{i}$, bulge-to-total ratios of both $M_{*}$ and light in the $g, r$ and $i$ bands and stellar and baryonic specific angular momenta of both the overall galaxy and its disc component alone.
Of these, $\Delta\rm{MS}$, $C_{i}$ and $j_{*}$, which trace the degree of star-formation activity, stellar structure and stellar kinematics respectively, show the strongest trends with the scatter.
Therefore, hereafter we focus on these three parameters and show related results for the xGASS centrals matched to the \citetalias{Yang2007} group catalogue in Fig. \ref{fig:HIHMscatter_Yang}.
The left column shows the \HI-halo mass parameter space with points colour-coded by either $\Delta\rm{MS}$ (panel a), $C_{i}$ (panel c) or $j_{*}$ (panel e).
The corresponding right column shows the correlation of these galaxy properties with the central's deviation from the median model relation ($\Delta M_{\rm{HI}}$), where the points are colour-coded by their respective $M_{\rm{halo}}$ values.
Note that we repeat our analysis from this section on the xGASS centrals matched to the \citetalias{Saulder2016} group catalogue and find consistent results (see Fig. \ref{fig:HIHMscatter_Saulder} for details).
Since our results do not vary significantly when using the different group catalogues, hereafter we focus on the results from using the \citetalias{Yang2007} group catalogue and present equivalent figures from using the \citetalias{Saulder2016} group catalogue in appendix \ref{sec:SaulderFig}.

\subsubsection{Position relative to the star-forming main-sequence}
\label{sec:dMSscatter}

Given the link between \HI\ and star formation, we can expect to find that the position of centrals in the \HI-halo mass parameter space correlates with their position relative to the SFMS.
The first row of Fig. \ref{fig:HIHMscatter_Yang} shows that, at a fixed $M_{\rm{halo}}$, higher $M_{\rm{HI,c}}$ corresponds to higher $\Delta\rm{MS}$ and vice versa.
This means that, not surprisingly, the scatter of the \HI-halo mass relation varies with $\Delta\rm{MS}$.
This is quantitatively shown in the right panel, where a $\rho_{\rm{S}} = 0.53$ indicates a significant correlation between $\Delta M_{\rm{HI}}$ and $\Delta\rm{MS}$.

Interestingly, in Fig. \ref{fig:HIHMscatter_Yang}b, we see a secondary dependence of the scatter on $M_{\rm{halo}}$, where $\Delta M_{\rm{HI}}$ values are higher for centrals in host halos of higher $M_{\rm{halo}}$ and vice versa.
This is because star-forming centrals have a significant positive \HI-halo mass correlation ($\rho_{\rm{S}} = 0.54$) and, while they dominate at the low $M_{\rm{halo}}$ end ($\rm{log}_{10}$$(M_{\rm{halo}}/\rm{M}_{\odot}) < 12.7$), their importance in driving the median trend decreases with increasing $M_{\rm{halo}}$.
In other words, star-forming centrals move away from the flat median \HI-halo mass relation as $M_{\rm{halo}}$ increases.
Conversely, the passive centrals represent the typical galaxy ($\Delta M_{\rm{HI}} = 0$) at the higher $M_{\rm{halo}}$ end, but lie significantly below the median at lower $M_{\rm{halo}}$ values.
This suggests that different populations drive the \HI-halo mass relation at different $M_{\rm{halo}}$, indicating that just computing medians might be reductive and not very informative, as in the case of the SFR-$M_{*}$ relation. 
We expand on this point in section \ref{sec:Discussion}.

While a high correlation between $\Delta\rm{MS}$ and $\Delta M_{\rm{HI}}$ is found, this cannot be interpreted as SFR driving the \HI\ mass in centrals. 
Indeed, if anything, the amount of cold gas in a galaxy regulates the amount of star formation \citep{Schmidt1959,Schmidt1963,Kennicutt1989,Kennicutt1998}.
Thus, in panels c-f, we investigate other galaxy properties that may be driving the scatter of the \HI-halo mass relation.

\subsubsection{Stellar-concentration index}
\label{sec:morph}

It is largely understood that the majority of \HI\ lies in the discs of galaxies \citep[\eg][]{Kerr1977,Knapp1987} as the rotational support in discs allows the gas to be stable against collapse.
While direct measurements of galaxy kinematics are difficult to acquire for a large sample of galaxies, proxies for galaxy structure are more readily available.
In this subsection, we look at one such proxy of structure: light-concentration in the $r$ band.

The second row of Fig. \ref{fig:HIHMscatter_Yang} shows that the scatter follows similar trends with $C_{i}$ as it does with $\Delta\rm{MS}$, but with a weaker correlation, $\rho_{\rm{S}} = -0.32$.
At a fixed $M_{\rm{halo}}$, the disc-dominated centrals tend to be more \HI-rich than the high stellar-concentration centrals.
The secondary dependence on $M_{\rm{halo}}$ is also seen here, with high stellar-concentration centrals lying below the relation at low $M_{\rm{halo}}$ values, but on the relation in the highest $M_{\rm{halo}}$ bin.
If we focus only on the disc-dominated centrals, $M_{\rm{HI,c}}$ and $M_{\rm{halo}}$ show a moderate correlation ($\rho_{\rm{S}} = 0.46$).

We repeat this analysis with the bulge-to-total ratios ($B/T$) presented in \citet{Cook2019} for xGASS galaxies.
However, we find that the majority of \HI-detected centrals are pure discs and assigned $B/T = 0$ and those centrals with $B/T > 0$ are dominated by $M_{\rm{HI,c}}$ upper limits.
This low dynamic range of $B/T$ for pure-discs in combination with observational limitations towards bulge-dominated systems makes it difficult to study the scatter's dependence on $B/T$.
In comparison, $C_{i}$ has a higher dynamic range for pure-discs, depending on the compactness of their stellar component, being an estimate of the steepness of their light profiles.
However, $C_{i}$ is not robust against all biases \citep[\eg\ edge-on spirals can be mistaken for early-type systems, see][]{Strateva2001,Masters2010} and only has a weak correlation with $\Delta M_{\rm{HI}}$.

\subsubsection{Stellar specific angular momentum}
\label{sec:jScatter}

In this subsection, we explore the kinematic state of the subset of 427 xGASS centrals with values of $j_{*}$ available. 
Overall, the trends are the same as those found with $\Delta\rm{MS}$ and $C_{i}$, but the correlations appear somewhat stronger, as shown in the bottom row of Fig. \ref{fig:HIHMscatter_Yang}. 
$j_{*}$ and $\Delta M_{\rm{HI}}$ have a high correlation coefficient, $\rho_{\rm{S}} = 0.55$, and the secondary dependence on $M_{\rm{halo}}$ appears more clearly defined: \ie\ $M_{\rm{HI,c}}$ of the $j_{*}$ sub-sample increases with $M_{\rm{halo}}$, gradually deviating further from the flat median \HI-halo mass relation.

It is important to note that we focus on $j_{*}$ in this study, even though $\Delta M_{\rm{HI}}$ shows a significantly higher correlation with baryonic specific angular momentum \citep[$j_{\rm{b}}$; provided by][]{Hardwick2022a}.
The way \citet{Hardwick2022a} estimated $j_{\rm{b}}$, makes it strongly correlated with $M_{\rm{HI}}$ ($\rho_{\rm{S}} = 0.89$) by construction, so any correlation may just be indicative of the bias in its estimation method. 
Conversely, $j_{*}$ has no built-in correlation with $M_{\rm{HI}}$.
Thus, $j_{*}$ is a more reliable avenue to investigate the role of specific angular momentum in driving the scatter of the \HI-halo mass relation.

Lastly, before concluding that specific angular momentum may be the primary driver of the scatter of the \HI-halo mass relation, we note that the analysis based on $j_{*}$ only uses a sub-sample of the all the xGASS centrals, limited to \HI-detected galaxies of $i > 30^{\circ}$. 
Therefore, we must re-check the correlation of the scatter with the three galaxy properties tested in this section, using the same sub-sample of centrals (\ie\ those catalogued with both $\Delta \rm{MS}$ and $j_{*}$ values).
Doing so, we find the correlation of $\Delta M_{\rm{HI}}$ is $\rho_{\rm{S}} = 0.55 (0.50)$ with $j_{*}$, $\rho_{\rm{S}} = 0.25 (0.21)$ with $\Delta \rm{MS}$ and insignificant ($\rho_{\rm{S}} = -0.12$) with $C_{i}$, for the selected sub-sample of centrals matched to the \citetalias{Yang2007} (\citetalias{Saulder2016}) group catalogue.
Thus, we conclude that the scatter of the \HI-halo mass relation shows the strongest correlation with specific angular momentum.

\section{Discussion}
\label{sec:Discussion}
\subsection{The median \HI-halo mass relation}

\begin{figure*}
	\includegraphics[width=0.9\textwidth]{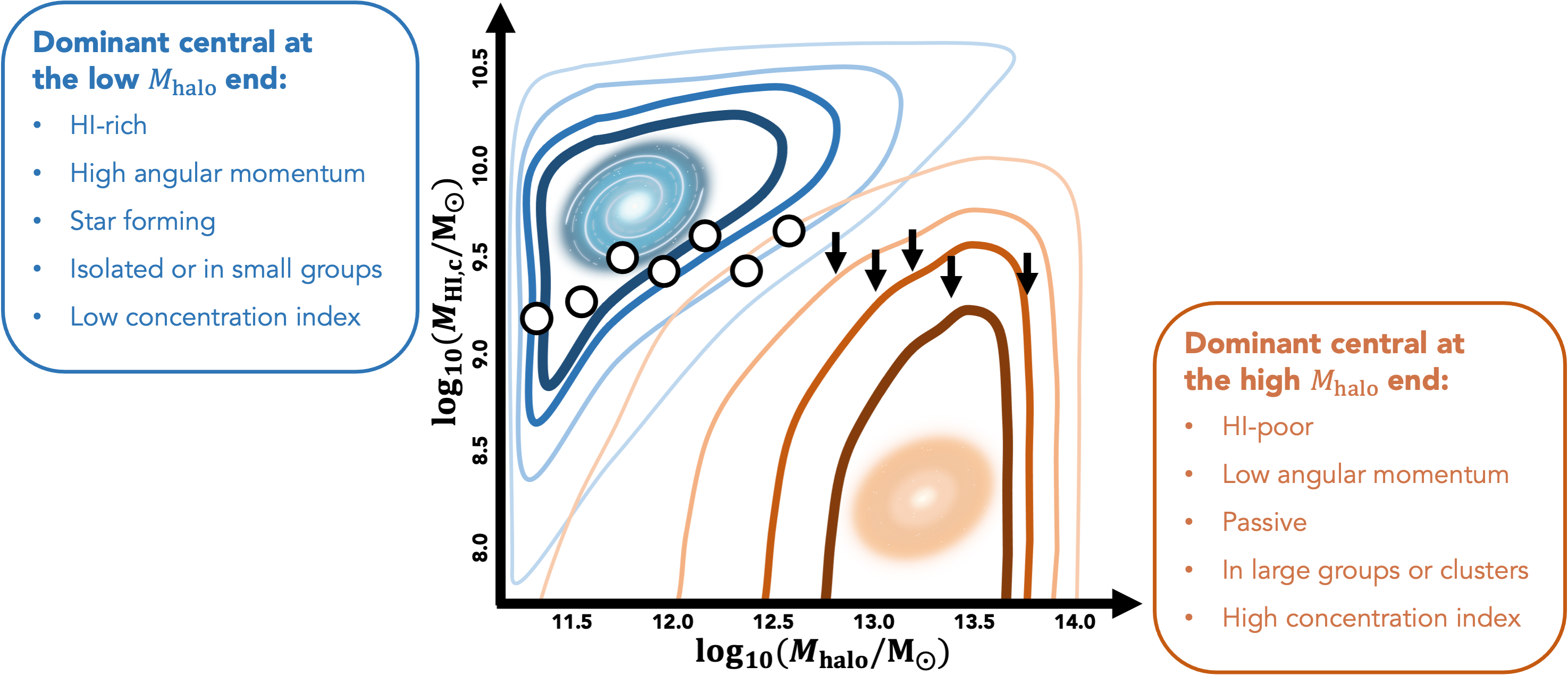}
    \caption{Diagram summarising the drivers of a flat median \HI-halo mass relation of centrals.}
    \label{fig:summary}
\end{figure*}

The median \HI-halo mass relation of xGASS centrals is flat at $\rm{log}_{10}$$(M_{\rm{HI,c}}/\rm{M}_{\odot}) \approx 9.40$.
For $\rm{log}_{10}$$(M_{\rm{halo}}/\rm{M}_{\odot}) < 12.7$, the median is a rotation-supported, star-forming and disc-dominated central, which tends to be \HI-detected and isolated. 
At higher $M_{\rm{halo}}$ values, the median is a low angular momentum, passive and high stellar-concentration central, which tends to be \HI-undetected and lies in groups ($N_{\rm{gal}} > 1$).
Therefore, at the low $M_{\rm{halo}}$ end, the \HI-rich centrals are dominant and drive the medians up whereas, at the high $M_{\rm{halo}}$ end, the \HI-poor centrals are dominant and drive the medians down (to the detection limit).
This switch in dominance of centrals of different properties leads to the flat median \HI-halo mass relation when median $M_{\rm{HI,c}}$ values are calculated across the entire sample of xGASS centrals, without binning by their secondary properties (see Fig. \ref{fig:summary} for visual aid).
This dependence of a scaling relation's slope on the underlying galaxy bi-modality is also seen in the \HI\ gas fraction-$M_{*}$ relation, where the slope arises because two different galaxy populations dominate at the two $M_{*}$ ends \citep{Brown2015}.

\begin{figure}
	\includegraphics[width=\columnwidth]{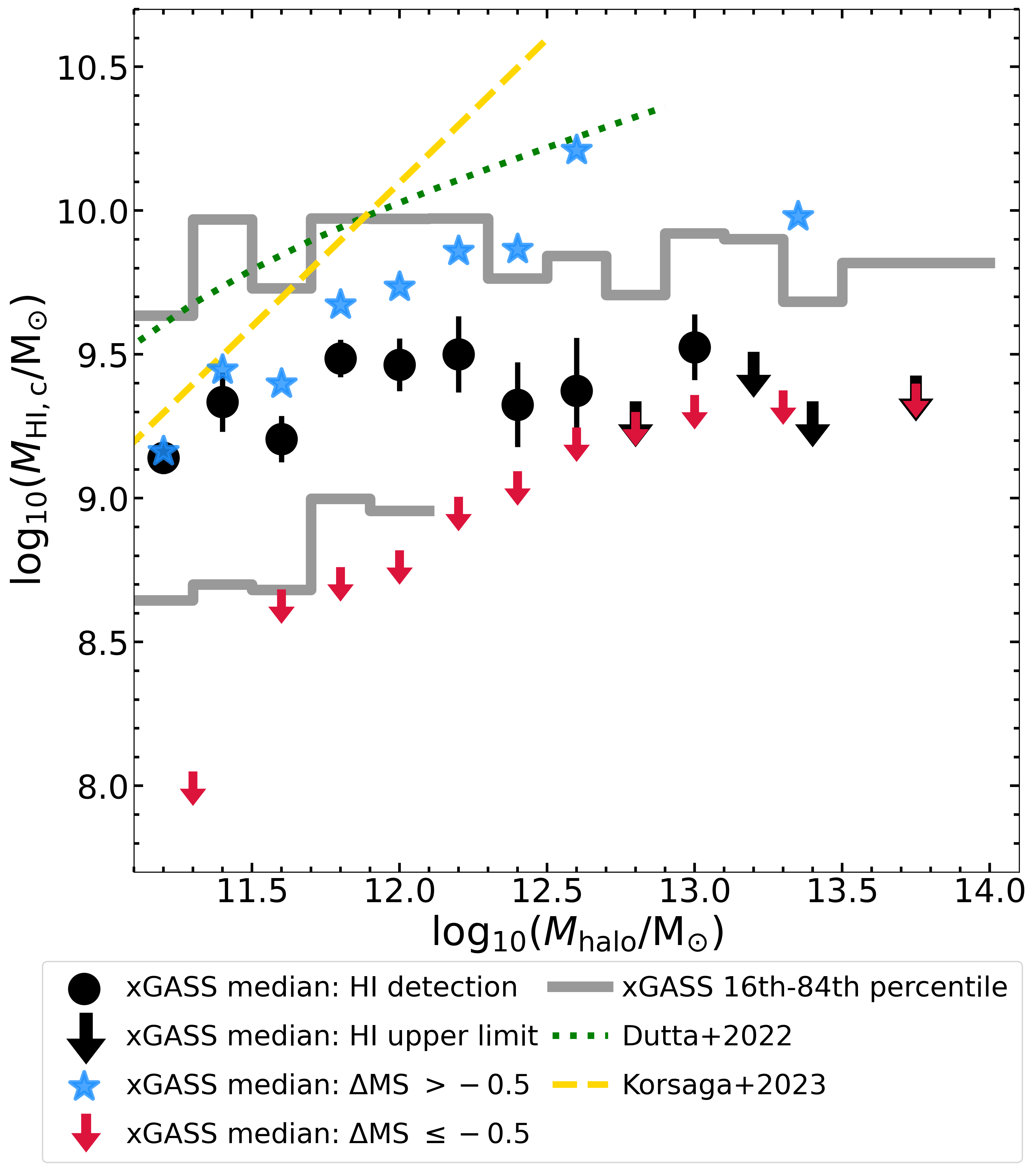}
    \caption{Comparing the \HI-halo mass relations presented in \citet{Dutta2022} (dotted green line) and \citet{Korsaga2023} (dashed yellow line) to the medians of xGASS centrals matched to the \citetalias{Yang2007} group catalogue.
    The xGASS median \HI\ mass values and 16th-84th percentiles are displayed as in Fig. \ref{fig:HIHM}a.
    The xGASS medians of star-forming centrals are shown as blue stars and those of passive centrals are shown as red downward-facing arrows.
    Note, the bins used to find the median \HI\ mass values of star-forming or passive centrals are different from each other and those used for finding the overall median median \HI\ mass values to ensure all bins have $> 10$ centrals.}
    \label{fig:pubRelCompIso_Yang}
\end{figure}

A flat median relation is an incomplete representation of the distribution of centrals in the \HI-halo mass parameter space.
The star-forming xGASS centrals, which tend to be rotation-supported, disc-dominated and isolated, showcase a high and positive correlation ($\rho_{\rm{S}} > 0.5$) between $M_{\rm{HI,c}}$ and $M_{\rm{halo}}$, indicative of an increasing \HI-halo mass relation (shown in Fig.~\ref{fig:pubRelCompIso_Yang}).
These centrals also showcase a positive trend between $\Delta M_{\rm{HI}}$ and $M_{\rm{halo}}$, \ie\ as their host halo mass increases they move further away from the flat median relation.
This is reminiscent of the distribution of galaxies on the SFR-$M_{*}$ parameter space \citep[\eg][]{Brinchmann2004,Elbaz2007,Noeske2007,Salim2007,Peng2010}, where there are multiple galaxy populations and, if we fit a median trend across all galaxies, we would not be able to isolate the SFMS from the passive population \citep[\eg][]{Saintonge2017,Catinella2018,Janowiecki2020}.

In a recent study, \citet{Korsaga2023} looked at the \HI\ masses of 150 isolated galaxies with extended discs and $i > 30 ^{\circ}$ from the Spitzer Photometry and Accurate Rotation Curves \citep[SPARC;][]{Lelli2016} survey and Local Irregulars That Trace Luminosity Extremes, The \HI\ Nearby Galaxy Survey \citep[LITTLE T\HI NGS;][]{Hunter2012} and measured their host halo masses between $9.4 < \rm{log}_{10}$$(M_{\rm{halo}}/\rm{M}_{\odot}) < 12.5$ from rotation curves.
For galaxies spanning 4 orders of magnitude in $M_{*}$, they found an increasing \HI-halo mass relation and a universal \HI-halo mass ratio, indicative of possible mass-independent self-regulation mechanisms.
In another recent study, \citet{Dutta2022} studied \HI-detected galaxies from the SPARC sample and 40\% of the Arecibo Legacy Fast Arecibo L-band Feed Array \citep[ALFALFA;][]{Haynes2011} survey area.
\citet{Dutta2022} treated all galaxies as isolated and estimated their host halo masses, between $9.4 < \rm{log}_{10}$$(M_{\rm{halo}}/\rm{M}_{\odot}) < 12.9$, by abundance-matching the \HI\ mass function with an \HI-selected halo mass function.
Doing so, they too recovered an increasing \HI-halo mass scaling relation with the functional form of a double power-law.

In Fig. \ref{fig:pubRelCompIso_Yang} (Fig. \ref{fig:pubRelCompIso_Saulder}), we compare the median \HI-halo mass relation of xGASS centrals matched to the \citetalias{Yang2007} (\citetalias{Saulder2016}) group catalogue with the scaling relations presented in \citet{Dutta2022} and \citet{Korsaga2023}.
In comparison to the xGASS medians, both scaling relations have on average 0.6-0.7 dex higher $M_{\rm{HI,c}}$ values. 
If we restrict the xGASS sample to star-forming galaxies, we recover a similar positive slope but with a negative offset of 0.3-0.5 dex from these published scaling relations.
This is not surprising given that both \citet{Dutta2022} and \citet{Korsaga2023} used samples that are more biased towards \HI-rich galaxies than xGASS.
Thus, our results, while confirming that $M_{\rm{HI,c}}$ monotonically increases with $M_{\rm{halo}}$ for star-forming centrals, highlight that it is challenging to accurately quantify the shape of the \HI-halo mass relation using only \HI-selected samples.
In the next subsection, we compare our findings to those of other observational studies that use spectral stacking methods on optically-selected samples.

\subsection{Comparison with stacking results}

\begin{figure}
	\includegraphics[width=\columnwidth]{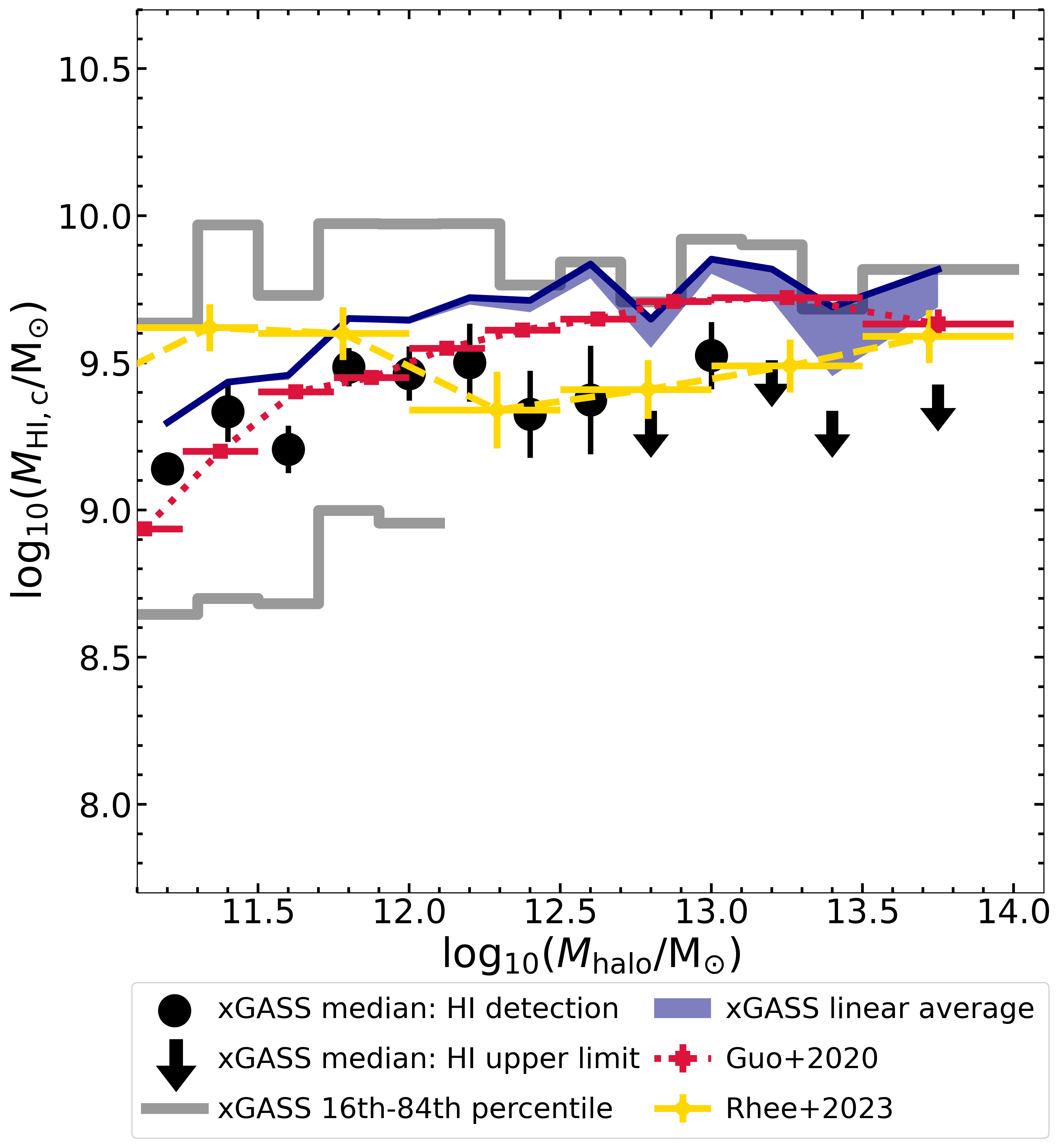}
    \caption{Comparing the stacked \HI-halo mass relations presented in \citet{Guo2020} (dashed yellow line) and \citet{Rhee2023} (dotted red line) to the linear averages of xGASS centrals matched to the \citetalias{Yang2007} group catalogue (dark blue shaded region).
    The solid dark blue line marks the highest linear averages.
    For reference, the xGASS median \HI\ mass values and 16th-84th percentiles are displayed as in Fig. \ref{fig:HIHM}a.}
    \label{fig:pubRelCompStacks_Yang}
\end{figure}

In Fig. \ref{fig:pubRelCompStacks_Yang} (Fig. \ref{fig:pubRelCompStacks_Saulder}), we compare the average \HI-halo mass relation of xGASS centrals matched to the \citetalias{Yang2007} (\citetalias{Saulder2016}) group catalogue with those found in spectral stacking studies of \citet{Guo2020} and \citet{Rhee2023}.
\citet{Guo2020} used the \citet{Lim2017} group catalogue to stack ALFALFA centrals, and measured the average $M_{\rm{HI,c}}$ values within a fixed aperture of diameter 200 kpc and velocity width 600 \kmps centred on the central.
\citet{Rhee2023} used the Galaxy And Mass Assembly (GAMA) group catalogue \citep{Driver2022} to stack centrals in the Deep Investigation of Neutral Gas Origins \citep[DINGO,][]{Meyer2009} survey, and measured the average $M_{\rm{HI,c}}$ values within a 49 kpc square aperture and 300 \kmps velocity width centred on the central.

Here, we work with linear averages instead of medians to make fair comparisons between the xGASS data and the stacked relations. 
Given that averages are more affected by non-detections than medians, we present two extremes for our average relation, estimated by setting non-detections to either 0 or their upper limit.
We shade between these lowest and highest averages and define this shaded area as the average \HI-halo mass relation of xGASS centrals.
As is evident, the \HI-undetected centrals have a limited impact on our average relation. 

The \HI-halo mass relation of centrals from \citet{Guo2020} agrees reasonably well with the xGASS averages, with an offset within 0.3 dex towards systematically lower $M_{\rm{HI,c}}$. 
The \HI-halo mass relation of centrals from \citet{Rhee2023} is significantly flatter than the xGASS averages.
The averages of \citet{Rhee2023} have a positive offset for $\rm{log}_{10}$$(M_{\rm{halo}}/\rm{M}_{\odot}) < 12$ because they only stack group centrals ($N_{\rm{gal}} \geq 2$) that, in this $M_{\rm{halo}}$ range, showcase flatter averages in both our sample and that of \citet{Guo2020}.
The averages of \citet{Rhee2023} have a negative offset for $12 < \rm{log}_{10}$$(M_{\rm{halo}}/\rm{M}_{\odot}) < 13.5$ because their sample does not include the \HI-rich isolated galaxies and, in this $M_{\rm{halo}}$ range, is more greatly dominated by the \HI-poor red centrals in comparison to the xGASS sample.
Curiously, the relation from \citet{Rhee2023} is in line with our median relation for $\rm{log}_{10}$$(M_{\rm{halo}}/\rm{M}_{\odot}) > 11.5$, however, we consider this to be just a coincidence given that it is meaningless to compare stacking results with median trends. 

\subsection{Comparison with simulations}

The scatter of the \HI-halo mass relation shows a correlation with $\Delta \rm{MS}$, $C_{i}$ and $j_{*}$ properties of centrals.
Of all these secondary properties, $\Delta M_{\rm{HI}}$ most strongly correlates ($\rho_{\rm{S}} \geq 0.5$) with specific angular momentum.
This strong correlation is consistent with the dependence between $M_{\rm{HI}}$ and specific angular momentum seen in baryonic Fall relations \citep[\eg][]{Murugeshan2020,ManceraPina2021a,ManceraPina2021,Hardwick2022a} and suggests that, at fixed $M_{\rm{halo}}$, the kinematic state of a central may be key in driving the amount of atomic gas available for star formation. 
This is also confirmed by the positive correlation we recover between $M_{\rm{HI,c}}$ and $M_{\rm{halo}}$ of rotation-supported and/or star-forming centrals.
Given the positive \HI-halo mass correlation of rotation-supported isolated galaxies and the potentially large scatter driven by passive group centrals, it is tempting to speculate that the scatter of the \HI-halo mass relation is driven by the host halo formation history.
This would be in line with the results from previous works that use (semi-)analytical models \citep[\eg][]{Guo2017,Chauhan2020}.
As such, it is interesting to see how our results compare with predictions from cosmological simulations. 

\begin{figure}
	\includegraphics[width=\columnwidth]{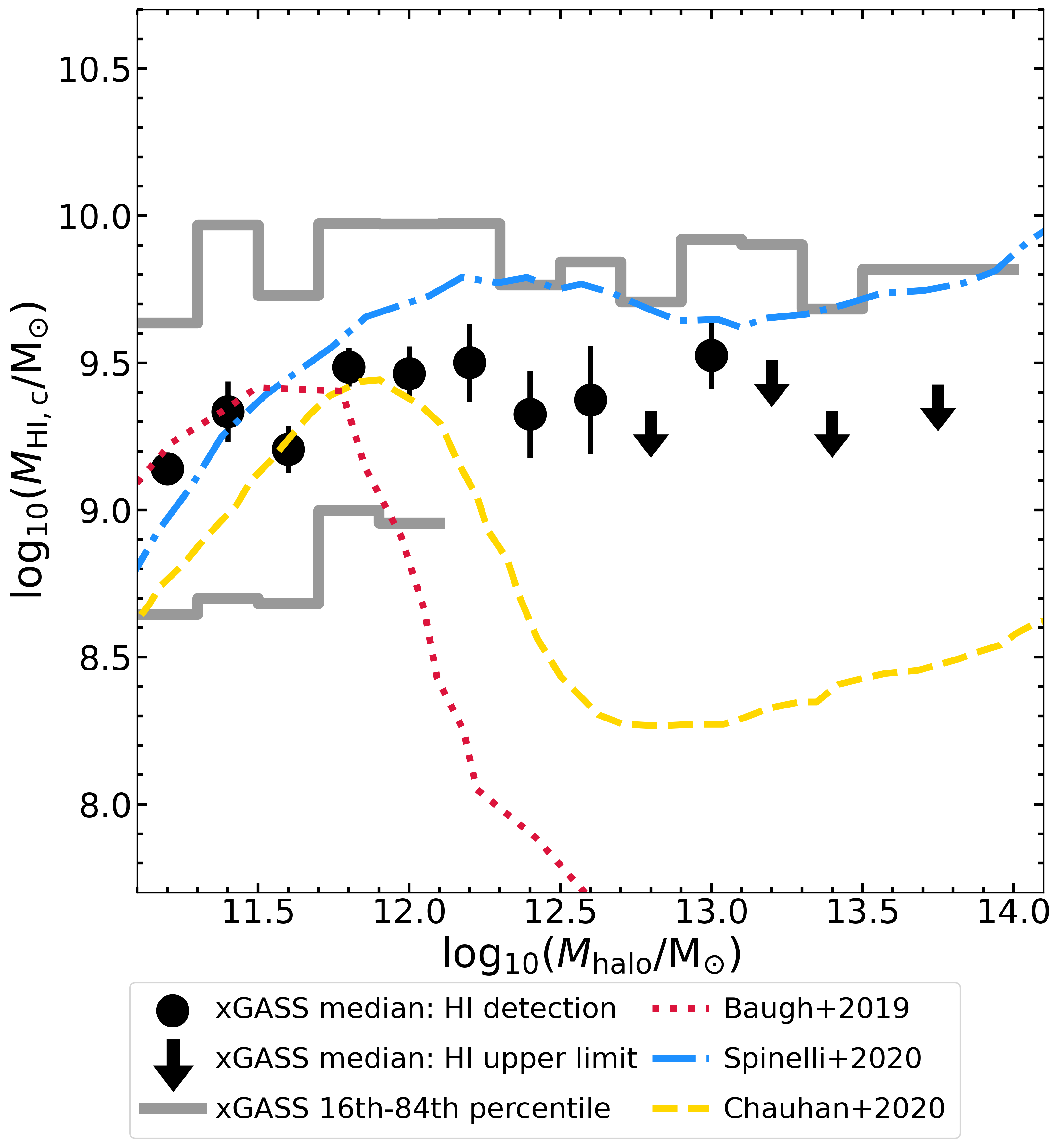}
    \caption{Comparing the median \HI-halo mass relations predicted from SAMs in \citet{Baugh2019} (dotted red line), \citet{Spinelli2020} (dash-dotted blue line) and \citet{Chauhan2020} (dashed yellow line) to the observed medians of xGASS centrals matched to the \citetalias{Yang2007} group catalogue.
    The xGASS median \HI\ mass values and 16th-84th percentiles are displayed as in Fig. \ref{fig:HIHM}a.}
    \label{fig:pubRelCompSimMed_Yang}
\end{figure}

\citet{Baugh2019}, \citet{Spinelli2020} and \citet{Chauhan2020} modelled the median \HI-halo mass relation of centrals using different semi-analytical models (SAMs) of galaxy formation: \texttt{GALFORM} \citep{Lacey2016}, GAlaxy Evolution and Assembly \citep[\texttt{GAEA},][]{DeLucia2014,Hirschmann2016,Xie2017,Zoldan2017} and \texttt{SHARK} \citep{Lagos2018} respectively.
In Fig. \ref{fig:pubRelCompSimMed_Yang} (Fig. \ref{fig:pubRelCompSimMed_Saulder}), we compare the medians of xGASS centrals matched to the \citetalias{Yang2007} (\citetalias{Saulder2016}) group catalogue with the median relations from the three different SAMs.
Although the 16th-84th percentile scatter of \texttt{GALFORM} and \texttt{SHARK} centrals is not published, from comparing the medians it appears that \texttt{GALFORM} and \texttt{SHARK} predominantly produce centrals that are populated with less \HI\ than that observed with xGASS between $11.4 < \rm{log}_{10}$$(M_{\rm{halo}}/\rm{M}_{\odot}) < 12.7$ and $11.9 < \rm{log}_{10}$$(M_{\rm{halo}}/\rm{M}_{\odot}) < 12.7$ respectively.
This is most likely due to a prescription of the AGN feedback that removes too much gas in large halos.
On the other hand, \texttt{GAEA} is more in line with the xGASS results, with consistent scatter and a systematic offset of $\sim 0.4$ dex towards higher median $M_{\rm{HI,c}}$ \citep[see Fig. B1 in][for details of the 16th-84th percentile scatter of \texttt{GAEA} centrals]{Spinelli2020}.
This demonstrates the potential of xGASS to provide tighter constraints to models, with particular emphasis on AGN feedback prescriptions.

It is important to acknowledge that some of the differences between xGASS and the predictions of theoretical models may also stem from the differences in the definition of central galaxies, combined with the challenge associated with accurately measuring halo masses from observations. 
Our use of two independent group catalogues offers some insights into these observational uncertainties (see Fig. \ref{fig:pubRelCompSimMed_Yang} and Fig. \ref{fig:pubRelCompSimMed_Saulder}). 
Moreover, \citet{Chauhan2021} show how the variations in the definition of centrals may impact the exact quantification of $M_{\rm{HI,c}}$ at fixed $M_{\rm{halo}}$.
However, it is extremely unlikely that the substantial differences between the results from xGASS and predictions from \texttt{GALFORM}/\texttt{SHARK} can be solely attributed to the disparity in the definition of centrals. 
Of course, the situation is different if a simulation estimates $M_{\rm{HI,c}}$ by taking into account the \HI\ associated with the entire central sub-halo, as done for example in \citet{Villaescusa-Navarro2018} with Illustris\texttt{TNG}, a cosmological magneto-hydro-dynamical simulation. 
In that case, a comparison with xGASS is completely meaningless and indeed the Illustris\texttt{TNG} relation (not shown) is systematically offset towards higher \HI\ masses.

Besides studying the shape of the \HI-halo mass relation of centrals, \citet{Chauhan2020} also investigated the physical properties driving its scatter.
They find a significant scatter, which is linked to the halo spin parameter ($\lambda$) for $\rm{log}_{10}$$(M_{\rm{halo}}/\rm{M}_{\odot}) < 11.5$.
In their `transition region', $11.5 < \rm{log}_{10}$$(M_{\rm{halo}}/\rm{M}_{\odot}) < 12.5$, the scatter depends on both $\lambda$ and the black hole-to-stellar mass ratio of the central galaxy ($M_{\rm{BH,c}}/M_{*,c}$): \ie\ centrals in host halos of high $\lambda$ and with low $M_{\rm{BH,c}}/M_{*,c}$ are \HI-rich and centrals in host halos of low $\lambda$ and with high $M_{\rm{BH,c}}/M_{*,c}$ are \HI-poor \citep[for details see Fig. 10 and 11 of][]{Chauhan2020}.
Finally, for $\rm{log}_{10}$$(M_{\rm{halo}}/\rm{M}_{\odot}) > 12.5$, the high $M_{\rm{BH,c}}/M_{*,c}$ washes out the effect of $\lambda$.
Overall, they find that centrals in host halos of high $\lambda$ have an increasing \HI-halo mass relation where $\rm{log}_{10}$$(M_{\rm{halo}}/\rm{M}_{\odot}) < 12.5$.
While $\lambda$ and the specific angular momentum of a central can be very different in reality, \texttt{SHARK} assumes that the dark matter and gas follow the same specific angular momentum and that the central and halo have aligned specific angular momenta.
Therefore, in low mass halos, where the main contribution of \HI\ is from the centrals, \citet{Chauhan2020} found the relation between $M_{\rm{HI}}$ and $\lambda$ to be a reflection of that between $M_{\rm{HI}}$ and the central's specific angular momentum.
Despite the mismatch in median relations, these findings from \citet{Chauhan2020} are qualitatively consistent with our results that, for centrals, the scatter of the \HI-halo mass relation has a strong correlation with specific angular momentum and that, for rotation-supported centrals, the $M_{\rm{HI,c}}$ increases with $M_{\rm{halo}}$.
However, our study cannot test if the correlation with specific angular momentum washes out at high $M_{\rm{halo}}$ due to high $M_{\rm{BH,c}}/M_{*,c}$, since we do not have measurements of $j_{*}$ for \HI-undetected centrals, which dominate at the high $M_{\rm{halo}}$ end.

\section{Conclusions}
\label{sec:End}
We use the xGASS $M_{\rm{HI}}$ catalogue and the \citetalias{Yang2007} group catalogue to study the scatter of the \HI-halo mass relation of 764 centrals.
Our derived median \HI-halo mass relation is flat at $\rm{log}_{10}$$(M_{\rm{HI,c}}/\rm{M}_{\odot}) \approx 9.40$ across halo masses, $11.1 < \rm{log}_{10}$$(M_{\rm{halo}}/\rm{M_{\odot}}) < 14.1$, consistent with the low correlation, $\rho_{\rm{S}} < 0.3$, found between $M_{\rm{HI,c}}$ and $M_{\rm{halo}}$.
We find that this flat \HI-halo mass relation arises from the combination of two different population of centrals dominating statistics at the opposite ends of the $M_{\rm{halo}}$ distribution.
Our medians are fully constrained by the data up to $\rm{log}_{10}$$(M_{\rm{halo}}/\rm{M}_{\odot}) < 12.7$, after which $M_{\rm{HI,c}}$ upper limits dominate the median relation.
Our results do not change significantly when, instead of the abundance-matching estimates from \citetalias{Yang2007}, we use dynamical $M_{\rm{halo}}$ estimates from \citetalias{Saulder2016} to conduct the same analysis.
This confirms that our study is robust against disagreements between different $M_{\rm{halo}}$ estimation methods.

We investigate the drivers of the scatter of the \HI-halo mass relation by studying secondary galaxy and halo properties.
We find a correlation of the scatter with the $\Delta\rm{MS}$, $C_{i}$, $j_{*}$ properties of the centrals.
The scatter most strongly correlates with $\Delta\rm{MS}$ and $j_{*}$, with a high correlation coefficient, $\rho_{\rm{S}} \geq 0.5$, between them and $\Delta M_{\rm{HI}}$.
Altogether, the typical galaxy represented by our flat median \HI-halo mass relation becomes more passive, centrally concentrated in stellar structure and dispersion-supported as $M_{\rm{halo}}$ increases.
These flat medians fail to showcase that the star-forming, disc-dominated and rotation-supported isolated galaxies in host halos of high masses tend to be \HI-rich.
This is reminiscent of the distribution of galaxies on the SFR-$M_{*}$ plane, suggesting that simple median or average relations may not be a complete representation of all centrals in the \HI-halo mass parameter space, especially at the high $M_{\rm{halo}}$ end.

When making comparisons with previous observation-based studies that use spectral stacking techniques, we find consistent results as the average \HI-halo mass relations found by \citet{Guo2020}, but we see a small tension with the result of \citet{Rhee2023}, who found a flatter average relation.
To further understand the true scatter while taking advantage of the improved number statistics enabled by the spectral stacking studies, we advise that centrals are also binned by their secondary properties (related to their kinematic state) before their spectra are co-added.

When comparing xGASS with cosmological simulations, we find that \texttt{GALFORM} and \texttt{SHARK} semi-analytical models predict extremely \HI-poor centrals in host halos above $\rm{log}_{10}$$(M_{\rm{halo}}/\rm{M}_{\odot}) \sim 12.3$.
We speculate that this is primarily an issue with the AGN feedback prescription in these models.
Conversely, among the models tested in this paper, \texttt{GAEA} shows the least tension with our results, with a relatively mild systematic offset towards higher $M_{\rm{HI,c}}$ values ($\sim 0.4$ dex) across the entire range of $M_{\rm{halo}}$ values investigated here.

We note that our study is limited at the high $M_{\rm{halo}}$ end, $\rm{log}_{10}$$(M_{\rm{halo}}/\rm{M}_{\odot}) \geq 12.7$, where \HI-undetected centrals start dominating.
Our study can be improved with follow-up surveys or targeted searches, conducted with the new and state-of-the-art radio telescopes such as the Five hundred meter Aperture Spherical Telescope \citep[][]{Nan2011} and the Square Kilometre Array \citep{Dewdney2009} and its precursors \citep{Johnston2008,Jonas2009,Koribalski2020a,Rhee2023}.
Observations from such surveys would not only have better resolution, enabling studies that use rotation curves to empirically measure $M_{\rm{halo}}$ \citep[\eg][]{Korsaga2023} and galaxy kinematics \citep[\eg][]{Fall2018,ManceraPina2021a}, but also better sensitivity, improving the number statistics.
Targeted \HI\ searches may also be able to provide $M_{\rm{HI,c}}$ detections or upper limits of centrals in very high mass halos, \ie\ $\rm{log}_{10}$$(M_{\rm{halo}}/\rm{M}_{\odot}) > 13.5$, where our study has only one wide $M_{\rm{halo}}$ bin and beyond which only a handful of observations exist \citep[\eg][]{McNamara1990,Jaffe1990,Taylor1996,Taylor1999,Veron-Cetty2000,Struve2010,Saraf2023}.

%%%%%%%%%%%%%%%%%%%%%%%%%%%%%%%%%%%%%%%%%%%%%%%%%%

\section*{Acknowledgements}
We thank the referee for constructive comments that improved this manuscript.
We thank Robin H.W. Cook for providing estimates of bulge-to-total ratios for xGASS galaxies that were used in our preliminary analysis.
L.C. acknowledges support from the Australian Research Council Discovery Project and Future Fellowship funding schemes (FT180100066, DP210100337).
Parts of this research were conducted by the Australian Research Council Centre of Excellence for All Sky Astrophysics in 3 Dimensions (ASTRO 3D), through project number CE170100013.

%%%%%%%%%%%%%%%%%%%%%%%%%%%%%%%%%%%%%%%%%%%%%%%%%%
\section*{Data Availability}
The xGASS catalogues used in this work are publicly available at \url{xgass.icrar.org/data.html}. 
Bulge-disc decomposition's from \citet{Cook2019}, and the matched catalogue between xGASS and \citetalias{Saulder2016} with our assigned environmental codes can be made available upon direct request to the authors.

%%%%%%%%%%%%%%%%%%%%%%%%%%%%%%%%%%%%%%%%%%%%%%%%%%
\section*{Author Contribution Statement}
LC devised this project.
MS lead the data analysis and drafted the manuscript with advice from LC and OIW.
BC led the xGASS team that observed the data.
SJ provided the matched group catalogues.
JAH provided the specific angular momentum estimates.
MS and LC are responsible for the interpretation of the results into the discussion. 
All authors provided constructive comments that improved this manuscript.


\bibliographystyle{mnras}
\bibliography{Paper3/references.bib} % if your bibtex file is called example.bib

%%%%%%%%%%%%%%%%% APPENDICES %%%%%%%%%%%%%%%%%%%%%

\appendix
\section{Dynamical Halo Mass Estimates}
\label{sec:SaulderFig}

\begin{figure*}
	\includegraphics[width=0.9\textwidth]{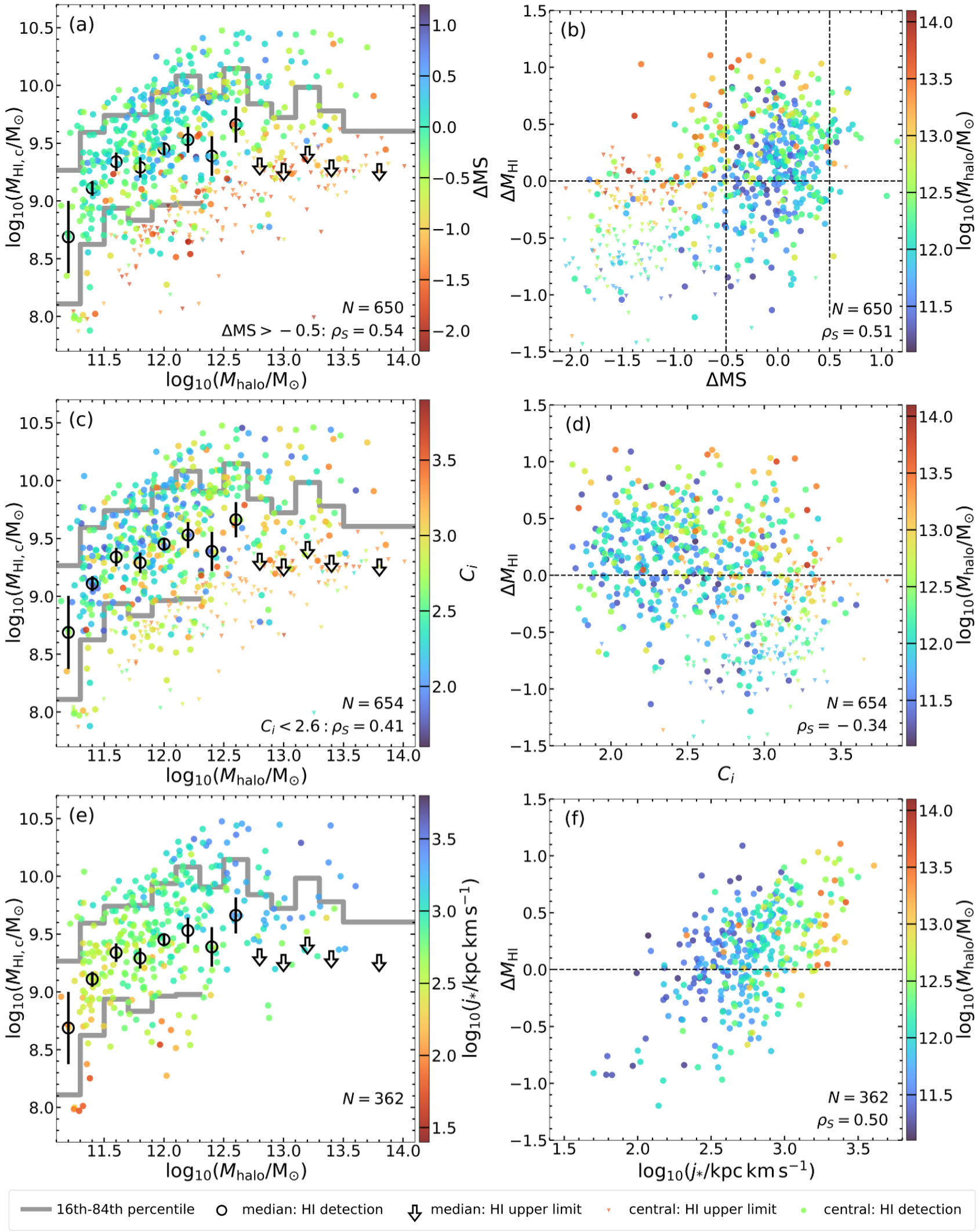}
    \caption{Same as Fig. \ref{fig:HIHMscatter_Yang}, but for the xGASS centrals matched to the \citetalias{Saulder2016} group catalogue, whose modelled median \HI-halo mass relation is given by equation \ref{eq:HIHMmodel_Saulder}.}
    \label{fig:HIHMscatter_Saulder}
\end{figure*}

\begin{figure}
	\includegraphics[width=\columnwidth]{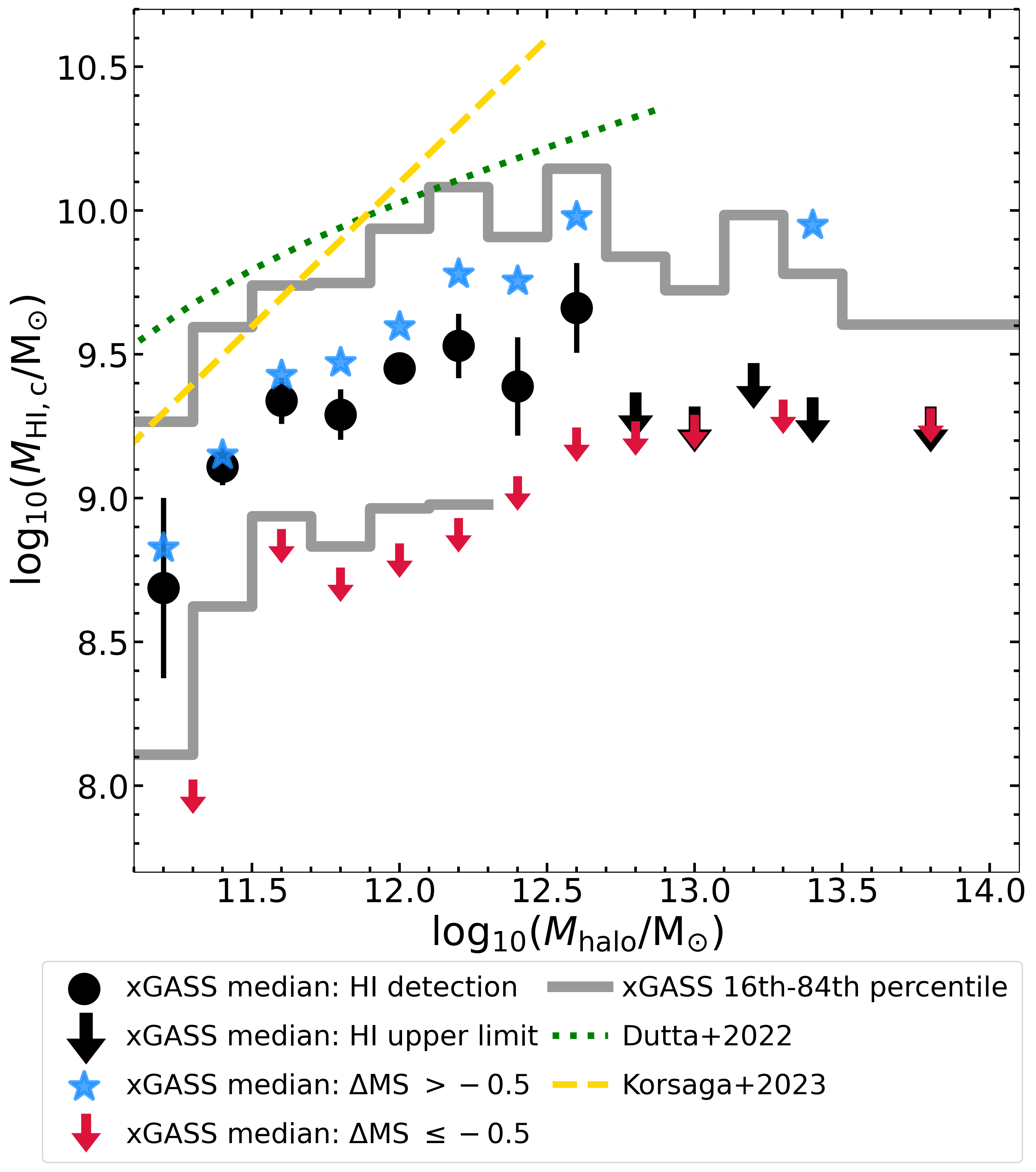}
    \caption{Same as Fig. \ref{fig:pubRelCompIso_Yang}, but for xGASS centrals matched to the \citetalias{Saulder2016} group catalogue.}
    \label{fig:pubRelCompIso_Saulder}
\end{figure}

\begin{figure}
	\includegraphics[width=\columnwidth]{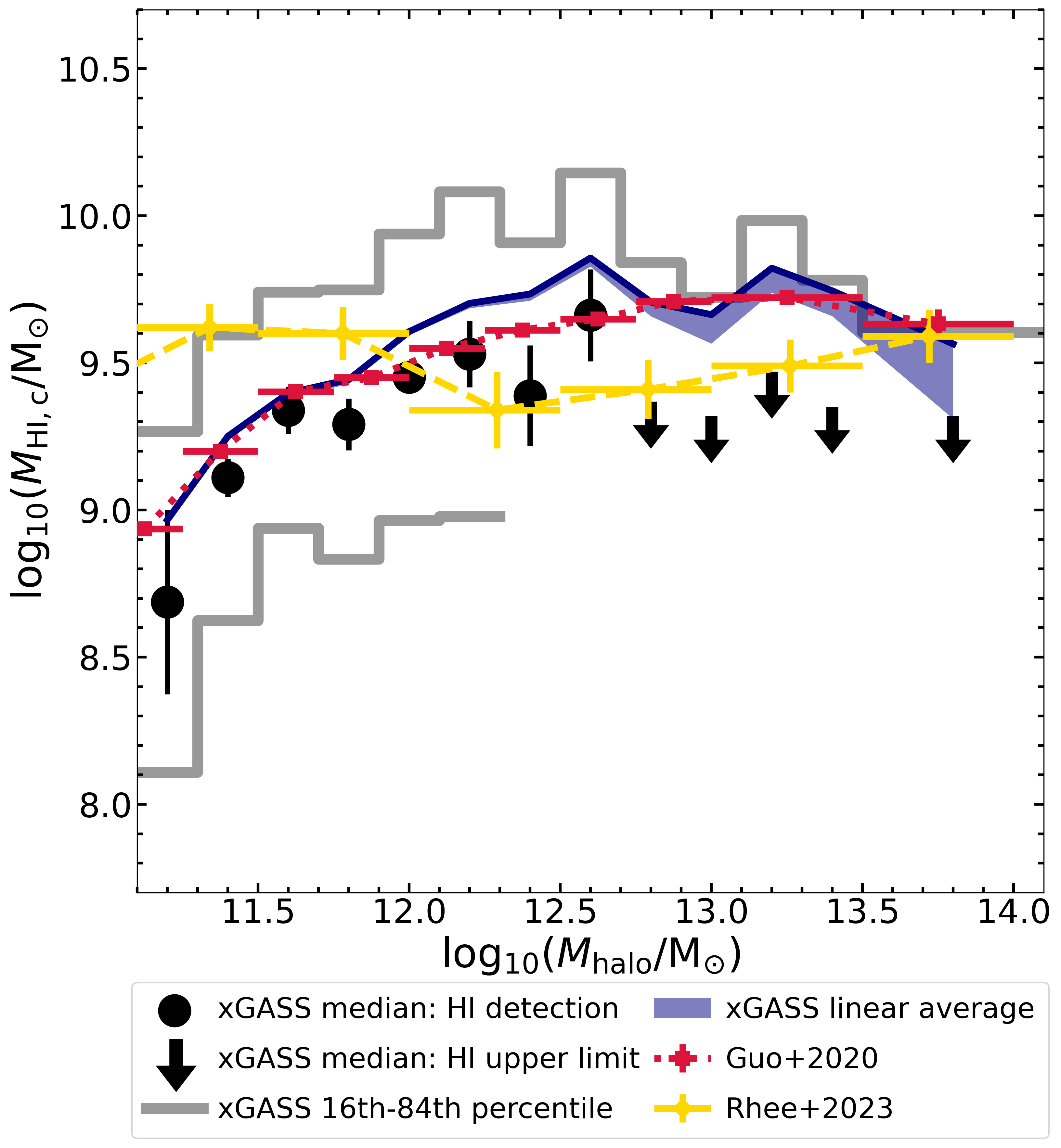}
    \caption{Same as Fig. \ref{fig:pubRelCompStacks_Yang}, but for xGASS centrals matched to the \citetalias{Saulder2016} group catalogue.}
    \label{fig:pubRelCompStacks_Saulder}
\end{figure}

\begin{figure}
	\includegraphics[width=\columnwidth]{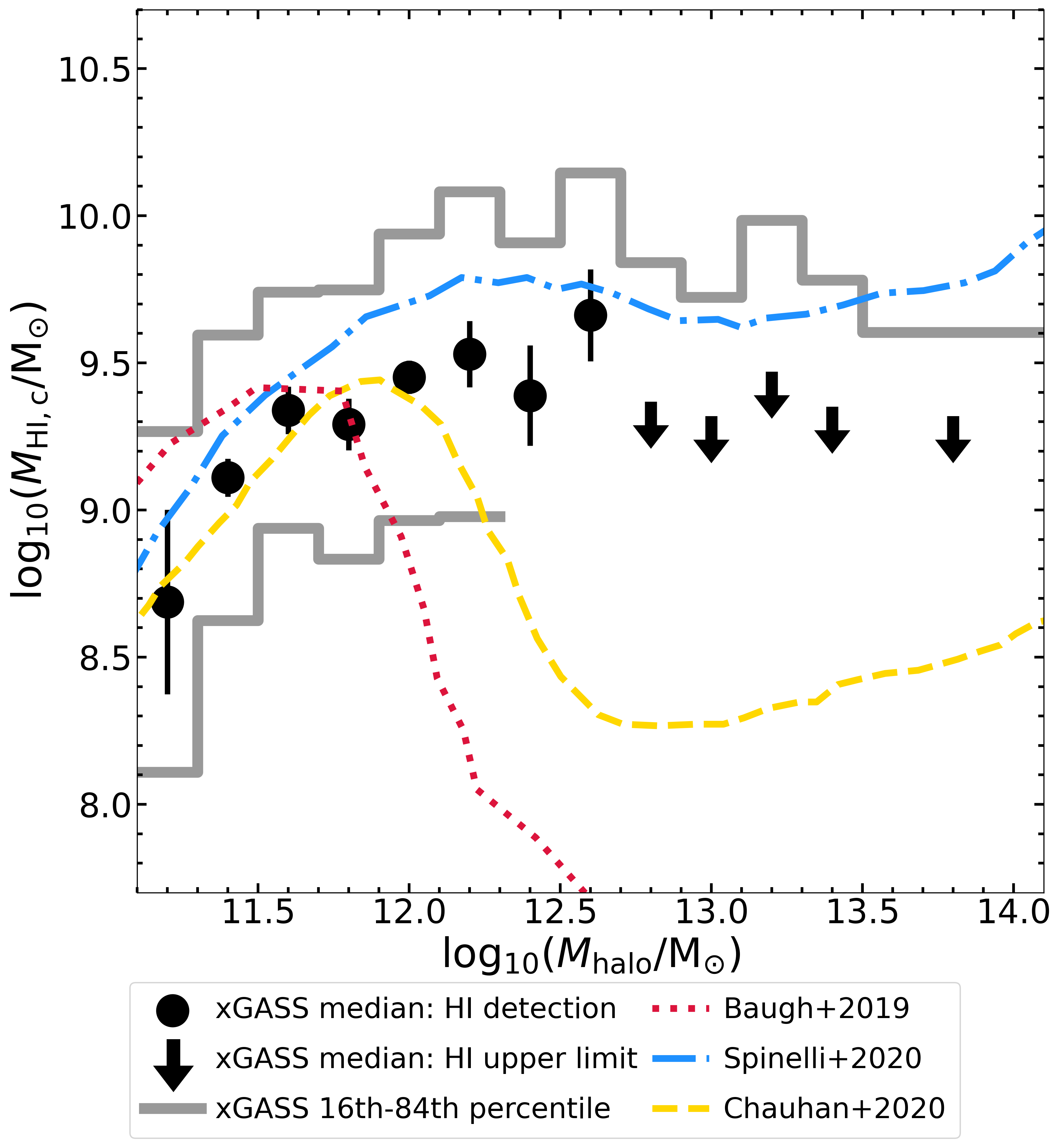}
    \caption{Same as Fig. \ref{fig:pubRelCompSimMed_Yang}, but for xGASS centrals matched to the \citetalias{Saulder2016} group catalogue.}
    \label{fig:pubRelCompSimMed_Saulder}
\end{figure}

Fig. \ref{fig:HIHMscatter_Saulder}, \ref{fig:pubRelCompIso_Saulder}, \ref{fig:pubRelCompStacks_Saulder} and \ref{fig:pubRelCompSimMed_Saulder} are the same as Fig. \ref{fig:HIHMscatter_Yang}, \ref{fig:pubRelCompIso_Yang}, \ref{fig:pubRelCompStacks_Yang} and \ref{fig:pubRelCompSimMed_Yang} respectively, but present the results for the xGASS centrals matched to the \citetalias{Saulder2016} group catalogue.
Overall, the results do not change significantly between the two group catalogues and, thus, are found to be robust against variances between the abundance-matching and dynamical $M_{\rm{halo}}$ estimation techniques.

%%%%%%%%%%%%%%%%%%%%%%%%%%%%%%%%%%%%%%%%%%%%%%%%%%

\bsp	% typesetting comment
\label{lastpage}
\end{document}